\documentclass[showpacs,preprintnumbers,amsmath,amssymb]{revtex4}
\usepackage{graphicx}
\usepackage{dcolumn}
\usepackage{bm}

\newcommand{\pdf}{\textsc{PDF}}
\begin{document}

\title{Self-Similarity in Decaying Two-Dimensional Stably Stratified Adjustment} 
\author
{Jai Sukhatme
 and Leslie M. Smith\\
   Mathematics Department, University of Wisconsin-Madison, Madison, WI 53706 \\}

\date{\today}

\begin{abstract} 

The evolution of large-scale density perturbations is studied in a stably stratified, two-dimensional 
flow governed by the Boussinesq equations. As is
known, intially smooth density (or temperature) profiles develop into fronts in the very early stages of evolution. 
This results in a 
frontally dominated $k^{-1}$ potential energy spectrum.
The fronts, initially characterized by a relatively simple geometry,
spontaneously develop into severely distorted
sheets that possess structure at very fine scales, and thus there is
a transfer of energy from large to small scales. It is shown here
that this process culminates in the establishment of 
a $k^{-5/3}$ kinetic energy spectrum, although its scaling extends over a shorter range 
as compared to the $k^{-1}$ scaling of 
the potential energy spectrum. The establishment of the kinetic energy scaling signals the onset of enstrophy decay 
which proceeds in a mildly
modulated exponential manner and possesses a novel self-similarity. Specifically, the self-similarity is seen in the time
invariant nature of the probability density function (\pdf{}) associated with the normalized vorticity field. Given the 
rapid decay of energy at this stage, 
the spectral scaling is transient and fades with the emergence of a smooth, 
large-scale, very slowly decaying, (almost)
vertically sheared horizontal mode with most of its energy in the potential component --- i.e.\ the Pearson-Linden regime. \\ 

\end{abstract}

\maketitle

\clearpage

\section{Introduction}

Large-scale geophysical flows usually evolve under the constraints of stable stratification and rotation. Indeed, it is known
that both of these constraints, individually and in concert, profoundly affect the motion of a fluid \cite{Gill}. Here we
restrict our attention to the stratified problem (see \cite{RL} for a review). In particular, we study the adjustment of a
two-dimensional stably stratified Boussinesq fluid to imposed large-scale density perturbations.  \\

Starting with the work of Riley, Metcalfe \& Weissman \cite{RMW}, there have been numerous studies of three-dimensional stably
stratified flows in both forced and decaying scenarios that indicate a spontaneous generation of layered structures from
initially isotropic fields (see for example \cite{Herring},\cite{Godeferd},\cite{SW}). Indeed, the small Froude number limit 
of the three-dimensional Boussinesq equations (especially in the decaying case) indicates layerwise motion with 
weak vertical correlation 
\cite{RMW},\cite{RL},\cite{Majda-book}. Further, amongst the variety of 
wave-vortex and wave-wave interactions 
possible in the three-dimensional case \cite{Godeferd},\cite{LR}, it appears as though 
the near resonant transfer into the slow 
(horizontal) wave 
modes via fast waves is responsible for the generation of these layered structures \cite{SW}. 
In fact the partitioning of energy between the fast and slow components in the general rotating
stratified system is an outstanding problem in atmospheric dynamics (for example Ford et.\ al \cite{Ford} and the references
therein). In light of this, it may be instructive to understand how energy re-distribution proceeds in the
relatively simplified, two-dimensional system that  
no longer possesses a distinct vortex mode and presents a 
situation wherein one can study the interaction of wave modes in isolation.  
A related issue is the fact that potential vorticity is identically zero
in two dimensions and a future goal is to understand if and how the
constraint of potential vorticity conservation in three-dimensional flows alters wave
interactions.\\

With regards to the two-dimensional problem, detailed numerical work was carried out by Bouruet-Aubertot, Sommeria \& Staquet
\cite{BS},\cite{BS1} wherein a bounded domain with no-flow conditions was considered. Focussing on the instabilities suffered by prescribed
flows, they showed that the wave-wave interactions result in a transfer of energy from large to small scales (see also Orlanski \&
Cerasoli \cite{Orlanski}).  Their analysis of imposed standing gravity waves indicated that these waves necessarily break after a
finite time (depending on the initial amplitude) and the spectral redistribution of energy proceeds via the so-called
wave-turbulence paradigm \cite{S-rev}. Here we lift the no flow condition by considering the problem in a periodic domain and
pursue the time evolution, from a state of rest, of a smooth large scale density (or temperature) perturbation. 
As one might anticipate, the aforementioned
wave-breaking is one stage in the overall scheme of things. \\

Apart from geophysical motivations, the system under consideration is part of an extended family of flows --- the so-called
dynamically {\it active scalars} \cite{Celani-active},\cite{Jai-chaos}.  In particular, statistical properties of the neutral and
unstably stratified Boussinesq systems have been the subject of recent investigations \cite{TS},\cite{Celani1},\cite{Celani2} and
the present work, wherein the system additionally supports waves, can be viewed as a continuation of these efforts. Further,
given the possibility of a finite-time breakdown of regularity in a two-dimensional setting, these problems are of considerable
mathematical interest and the reader inclined to pursue such matters is referred to Cordoba \& Fefferman \cite{Cordoba} (and the
references therein) for a recent overview. \\

The remainder of the paper is organized as follows : we first provide the basic equations and conservation laws along with the
setup of the numerical experiment.  Next we touch briefly upon the initial stages of evolution that have been well
documented in existing literature. We then proceed to the focus of the paper showing a 
$k^{-1}$ scaling for the potential energy spectrum followed by the establishment of a 
transient $k^{-5/3}$ scaling for the kinetic energy, the onset of nearly 
exponential decay of enstrophy and the
establishment of an invariant normalized vorticity \pdf{}. Due to the rapid decay of energy at this stage of the problem, the spectra are seen to
gradually shift downwards, their scaling becomes less distinct, and finally we see the emergence of a very slowly decaying, vertically
sheared, large-scale, predominantly horizontal mode --- most of whose energy is trapped in the potential component --- as elucidated 
by Pearson \& Linden \cite{Pearson}. 
We conclude by summarizing the
various stages, discuss the fate of stratified adjustment in the inviscid limit and point
out avenues for future research. \\

\section{The governing equations}

The equations governing two-dimensional stratified flow under the Boussinesq approximation are \cite{Majda-book} 

\begin{eqnarray}
\frac{D \vec{u}}{D t} = - \frac{\nabla p'}{\rho_0} + g \alpha \theta \hat{k} \nonumber \\
\frac{D \theta}{D t} + \lambda w = 0 ~;~ \lambda=\frac{B}{\alpha \rho_0} \nonumber \\
\frac{\partial u}{\partial x} + \frac{\partial w}{\partial z} = 0.
\label{1}
\end{eqnarray}
In the above $\rho = \rho_b + \rho', p = p_b + p'$ and $T=T_b+\theta$, where $\rho_b=\rho_0 + \hat{\rho}(z)$.  In particular,
$\hat{\rho}(z)=-Bz$ 
and the basic state 
is in hydrostatic balance, i.e.\  $\partial p_b/\partial z = - \rho_b g$. 
The temperature and density are
related via $\rho=\rho_o [ 1 + \alpha(T_0 - T)]$, and thus the basic state temperature profile is $T_b= T_0 + \hat{T}(z)$
with $\hat{T}(z)=-\hat{\rho}(z)/(\alpha \rho_o)$. System (\ref{1}) results from the assumption $\rho',\hat{\rho}(z) << \rho_0$.
As $\rho_o,\lambda,g,\alpha$ are constants we consider the perturbed fields $(\vec{u},p',\rho',\theta$)  in a periodic domain 
\footnote{Note that if we set $\lambda = \alpha g = 2 \Omega$, (\ref{1}) is equivalent to 
a (two component) three dimensional rotating flow that satisfies $\frac{\partial}{\partial y} = 0$ and 
has a rotation vector given by $(-2\Omega,0,0)$. }.
In vorticity-stream form (\ref{1}) reads

\begin{eqnarray}
\frac{D \omega}{D t} = -\alpha g \frac{\partial \theta}{\partial x} ~~;~~ \frac{D \theta}{D t} = -\lambda \frac{\partial \psi}{\partial x} \nonumber \\
\omega = - \nabla^2 \psi~;~ u=-\frac{\partial \psi}{\partial z}~;~ w=\frac{\partial \psi}{\partial x}.
\label{2}
\end{eqnarray}
Defining $E = \int_{\bf D} [u^2 + w^2 + \frac{\alpha g}{\lambda} \theta^2] $, we see that (\ref{1}) conserves 
$E$ --- which 
we refer to as the total energy of the system 
\footnote{There are some subtleties associated with the definition of the energy, more precisely 
the pseudoenergy --- see Shepherd \cite{Shepherd} for a discussion regarding this conservation law
and its related symmetry.}. \\ 

By linearizing (\ref{2}) it can be seen that the above supports gravity waves obeying the dispersion relation

\begin{equation}
\sigma(\vec{k}) = \pm \sqrt{(\lambda g \alpha)} \frac{k_x}{K}~;~ K^2 = k_x^2 + k_z^2.
\label{3}
\end{equation}
Note that the presence of a uniform gravitational field in (\ref{2}) breaks symmetry with respect to mirror images about the x-axis.  
Apart from this lack of complete parity invariance, 
in contrast to the usual two-dimensional scenario \cite{Kraichnan-Mont} the vorticity in (\ref{2}) is not a Lagrangian 
invariant. 
This indicates that the familiar two-dimensional inverse cascade of kinetic energy is unlikely to be realised as it 
relies, amongst other things, on the dual conservation of energy and enstrophy \cite{Kraichnan-Mont}. 
Also, in comparison to the free-convective case (i.e.\ $\lambda=0$) 
\cite{TS},\cite{Celani2}, functions of the form $f(\theta)$ are
not conserved by (\ref{2}). Hence an immediate direct cascade of $\theta^2$ is unlikely. 
Introducing dissipation via the usual kinematic viscosity and diffusivity 
(both taken to be $\gamma$) ---
i.e.\ $\gamma \nabla^2 \omega$ and $\gamma \nabla^2 \theta$ in (\ref{2}) --- we have

\begin{eqnarray}
\frac{\partial}{\partial t} \int_{\bf D} u^2 + w^2 = 2 \alpha g \int_{\bf D} \theta w -2 \gamma \int_{\bf D} \{|\nabla w|^2 + |\nabla u|^2 \} \nonumber \\
\frac{\alpha g}{\lambda} \frac{\partial}{\partial t} \int_{\bf D} \theta^2 = -2 \alpha g \int_{\bf D} \theta w -2 \gamma \frac{\alpha g}{\lambda} \int_{\bf D} |\nabla \theta|^2 \nonumber \\
\frac{\partial}{\partial t} \int_{\bf D} \omega^2 = - 2 \alpha g \int_{\bf D} \frac{\partial \theta}{\partial x} \omega -2 \gamma \int_{\bf D} |\nabla \omega|^2. \nonumber \\
\label{4}
\end{eqnarray}
So for $\gamma > 0$, $\frac{\partial E}{\partial t} < 0$ and (\ref{4}) imply that all fields have to die out in the 
limit $t \rightarrow \infty$.
Of course, even though $E$ is a monotonically decreasing function of time, the enstrophy (and higher moments of $\omega$) 
can show a significant increase 
before finally decaying away. \\

\section{The numerical experiment}

To study the process of adjustment, we start with an initial state at rest and in hydrostatic balance to which is imposed a large
scale $\theta$ perturbation. Various mathematical forms for the initial condition were used to check the qualitative similarity in the 
response of the system, in particular the results presented correspond to 
to a Gaussian bump ($\theta(x,z,0)= 0.5 \exp \{ -4 [(x-\pi)^2 + (z-\pi)^2] \}$) in the middle of the domain. 
System (\ref{2}) is solved at a resolution of $750 \times 750$ in a $2\pi$ periodic domain using a 
pseudo-spectral method de-aliased by the two-thirds rule and 
a fourth order Runge Kutta time stepping scheme. 
Defining
the Froude ($Fr$) and Reynolds ($Re$) numbers as $Fr=u/NL$ and $Re=u L/\gamma $ respectively, 
taking $N$ to be an $O(1)$ quantity,
consider an initial $\theta$ perturbation such that the total energy is also an $O(1)$ entity. 
Apriori we expect the following
stages of evolution : as the fluid starts from rest, initially both $Fr,Re \ll 1$ --- i.e.\  we have a laminar flow that is
strongly constrained by stratification.  Since $\gamma$ is small, as soon as a certain fraction of the potential energy is
converted to kinetic energy implying $\max{(u)} \sim 1$, we have $Fr \sim 1, Re \gg 1$ --- i.e.\ a turbulent flow that is weakly
constrained by stratification. Finally, because $\gamma > 0$ the fields must eventually decay so that we enter a diffusion and
stratification dominated regime. In accordance with this scenario we set the parameters in the present 
simulation to be $\rho_o=1, B=0.1, \alpha=0.1, \gamma=5 \times 10^{-4}$ and $N=1$.\\

\subsection{The formation of fronts and their subsequent distortion}

Fig. 1 shows the evolution of energy --- total, potential ($\int_{\bf D} \frac{\alpha g}{\lambda} \theta^2)$ and
kinetic ($\int_{\bf D} [u^2 + w^2]$) --- and enstrophy with time. Focussing on the early stages of development shown in the first
column of Fig. 2, as predicted from linear theory we see the generation of gravity waves that mediate the exchange of
energy between the potential and kinetic components. Indeed, simulations with differing initial conditions and variations in the
strength of stratification follow qualitatively similar paths though the quantitative partition of energy between the potential
and kinetic components is not identical in all situations. Further, at very early times ($t < 1.5$s), all gradient fields are quite
mild and the evolution is fairly inviscid. Along with this laminar evolution, Fig. 3 shows the
emergence of frontal structures perpendicular to the vertical direction. Moreover, this frontal development is accompanied by increasing
vertical shear and, even though this is not a steady flow, the accompanying decrease in Richardson number hints at the onset of
instability \cite{Phil_1}. In fact, as noted in previous studies such a situation necessarily leads to wave-breaking \cite{BS}. In physical
space we begin to see the fronts evolving into highly convoluted sheets and the wave-wave interactions result in a redistribution
energy from large to small scales \cite{BS},\cite{Orlanski}. 
Of course as small-scale structures are being created, we see that
the relatively inviscid behaviour seen for $t < 1.5$s ceases and dissipation of energy begins to increase. \\

\subsection{Enstrophy decay, invariant \pdf{}s and spectral scaling}

Proceeding to the focus of this communication, from Fig. 1 we notice a marked change in the behaviour of the enstrophy
as the fronts become severely distorted. Specifically the enstrophy, which grew during the initial front formation and subsequent
development, now decays in a fairly monotonic manner. Examining the enstrophy in detail (see the $\log$
plot in Fig. 4), immediately after the maximum we notice that the primary signature of the 
decay is exponential along with a secondary small
amplitude modulation.  In fact, Fig. 5 shows the vorticity field well after the 
enstrophy attains it maximum value (see
the figure captions for the exact times of the snapshots); as is expected via the non-conservation of vorticity from
(\ref{2}), we do not see successive mergers resulting in large-scale structures, but rather the vorticity field continues to
consist of distinct blobs separated by sharp ridges of concentrated enstrophy dissipation. \\

Motivated by a somewhat similar scenario in the decaying passive scalar problem \cite{Sinai-Yakhot},\cite{Jai-1},\cite{Jai-2} ---
where the decay (of the passive field) is purely exponential --- we consider the normalized 
variable $X=\omega/Q^{1/2}$, where $Q=<\omega^2>$ and
$<\cdot>$ denotes spatial averaging.  From (\ref{2}) and (\ref{4}), after performing a spatial average, the equation governing
$<X^{2n}>$ is

\begin{equation}
\frac{Q}{2n} \frac{\partial}{\partial t}< X^{2n} > = [\gamma Q_1 + \alpha g Q_2 ] < X^{2n} >
 - \alpha g Q^{\frac{1}{2}} < X^{2n-1} \frac{\partial \theta}{\partial x} > - (2n-1) \gamma Q < X^{2n-2}  (\nabla X)^2 >
\label{6}
\end{equation}
where $Q_1=<(\nabla \omega)^2>$ and $Q_2=<\omega \frac{\partial \theta}{\partial x}>$.
In the passive scalar problem, $X=\phi/Q_p^{1/2}$ where $\phi$ denotes the passive field and $Q_p=<\phi^2>$. In that case 
the purely exponential
decay of $<\phi^2>$ (and higher moments) led to 
$\frac{\partial <X^{2n}>}{\partial t} = 0$ and consequently $\pdf{}(X)$ 
attained an invariant profile \cite{Jai-1},\cite{Jai-2}. 
However in the present case, the moments in (\ref{6}) inherit the 
secondary modulation from 
$<\omega^2>$ and fluctuate about a mean. In fact, as all the moments have the same temporal fluctuations, the entire \pdf{}
is expected to attain an invariant shape but will exhibit small shifts in magnitude.
The extracted \pdf{}s --- see Fig. 6 --- are plotted in three groups. The upper panel focusses on 
early times ($t < 7$s) and shows the approach to self-similarity. Interestingly, this approach 
is characterized by a gradual decrease in intermittency, i.e.\ the \pdf{}s in the initial stages of evolution are extremely
fat tailed 
--- either stretched exponentials or power laws keeping in mind the difficulty in distinguishing between these two
functions \cite{Yakhot-etal},\cite{Jai-2}. \\

The middle panel of Fig. 6 shows the \pdf{}s for the interval $t \in [12,28]$s, i.e.\ the time
during which the enstrophy experiences a modulated exponential decay. The self-similarity is
evident, in fact the \pdf{}'s are now characterized by a small Gaussian core and an exponential tail. 
The $\theta$ field during this interval is shown in Fig. 7 --- 
notice that even though there exists a background layering, the frontal structures that ride on this background are 
oriented quite randomly. In other words, irrespective of its direction a one-dimensional cut of the 
snapshots 
in Fig. 7 will encounter a frontal jump.
In effect the picture
that emerges is, as the frontal structures become unstable ($t \approx 1.5$s), energy that was trapped at larger scales during the
very early stages of evolution begins to fill out the entire available range of scales. This "filling out" proceeds --- from $t
\approx 2-7$s --- until one achieves a state wherein the kinetic energy spectrum is close to a $k^{-5/3}$
scaling 
while the potential 
energy continues to scale as $k^{-1}$ (though at slightly larger scales) --- this can be seen in Fig. 8
which shows the kinetic and potential energy spectra from $t \approx 7$s onwards. Of course as is seen in Fig. 8,
the spectra gradually shift downwards as the total energy is decaying, the diffusive roll-off extends to larger
scales and the scaling becomes less distinct (especially in the kinetic energy) as the fields become 
progressively smoother. In spite of this change in the nature of the flow, the enstrophy continues to decay
in a modulated exponential manner and the self-similarity in the normalized 
vorticity field persits throughout the interval 
$t \in [12,28]$s. \\

Regarding the scaling of the energy spectra, 
the phenomenology of stably stratified turbulence 
suggests the presence of two regimes \cite{Phil_2} (see especially the discussion in Section 3 of Phillips \cite{Phil_2}), 
an inertial regime wherein the kinetic energy scales isotropically 
as $k^{-5/3}$ \cite{Lumley},\cite{Phil_2} and
a bouyancy regime wherein the scaling is anisotropic and goes as $k_z^{-3}$ \cite{BS1},\cite{Phil_2}. 
At the present stage, $Re >>1$  coupled with 
$Fr \sim O(1)$ implies a turbulent regime that is weakly constrained by bouyancy. This, along with the observed
directional independence of the spectra, suggests the applicability of inertial range phenomenology and thus 
supports the $k^{-5/3}$ scaling of the kinetic energy. 
Given the weak bouyancy constraint, the scaling of the potential energy spectrum is dictated by the geometry of the $\theta-$ field.
Specifically, as mentioned earlier even though the fronts become unstable
they retain much of their identity, i.e.\ instead of being smoothed out, they evolve into highly distorted 
sheets (Fig. 7).  In fact, these step-like features 
continue to dominate
the potential energy spectrum resulting in the $k^{-1}$ scaling \cite{Vassi}, \cite{Ray-fractal}. \\

\subsection{Vertically sheared horizontal flows (a Pearson-Linden regime)}

Returning to the energetics of the flow, Fig. 1 indicates that at long times, i.e.\ $t > 40$s, the total energy
reverts to an extremely slow decay. Further, in this stage almost all of the energy in the system is in the potential component
(see also the third column of Fig. 2). The minute amount of kinetic energy indicates that a dissipative linear
analysis would be appropriate --- in fact, we are precisely in the last stages of decaying stratified turbulence as elucidated by
Pearson \& Linden \cite{Pearson} (their analysis was more detailed with diffusivity $\neq$ viscosity).  Substituting a Fourier
decomposition into the dissipative linearized form of (\ref{2}), we have

\begin{equation}
\psi(\vec{x},t) = \int \hat{\psi}(\vec{k})~ \exp \{ i(k_xx+ k_zz) \} \exp \{-\gamma K^2 t \} \exp \{ \pm i \frac{k_x}{K} \sqrt{(\lambda \alpha g)} t \} ~d\vec{k}.
\label{7}
\end{equation}
The oscillatory nature of the exponential results in a dominance of the integral by modes with $k_x \approx 0$. Further, at long
times a subset of these modes with the smallest rate of decay will remain. 
Hence we are left with a flow wherein $w \approx 0$, $u = u(z)$ and
$\theta= \theta(z)$, i.e.\ smooth vertically sheared horizontal flows with vertical structure restricted to the smallest wavenumbers (largest scales) 
as is seen in Fig. 9
\footnote{Previous numerical work by Bouruet-Aubertot et. al.\ \cite{BS1} shows the potential and kinetic energy spectra to scale
as $k_z^{-3}$, i.e.\ their system appears to satisfy the bouyancy range phenomenology. Referring to Fig. 3 in \cite{BS1} 
the plotted time averaged spectra are seen to follow the $k_z^{-3}$ scaling at fairly large times. Further, they indicate that the decay of
energy is quite slow. Indeed, this is consistent with the
emergence of the Pearson-Linden regime which is anisotropic, decays slowly and is characterized by smooth fields, 
hence much steeper spectral slopes. In our
case, the Pearson-Linden regime is very smooth and does not show a power law scaling. But as our diffusivity is greater than the 
one used in Bouruet-Aubertot et. al.\ \cite{BS1}, it is possible that reducing $\gamma$ will yield a Pearson-Linden regime
that shows the bouyancy range scaling. Unfortunately, given the presence of strong fronts, to keep the spectral code well posed 
reducing $\gamma$ involves the use 
of a hyper-viscocity 
or a high wavenumber filter. At present we have chosen not to employ such smoothening techniques and are limited to 
$\gamma \sim 5 \times 10^{-4}$. }. 
Regarding the vorticity field, from Fig. 4 we see that the nature of the enstrophy decay changes as we proceed
into the Pearson-Linden regime --- indeed, the set of \pdf{}'s in the lowermost panel of Fig. 6 correspond to 
these late times and as would be expected from a diffusive regime, the \pdf{} gradually acquires more of a Gaussian form. \\

Note that the form of (\ref{7}) does not indicate how energy is transferred into 
the vertically sheared horizontal modes. Indeed a study of the forced 2D case \cite{Smith-short} suggests an 
inverse tranfer of energy into these quasi-horizontal modes, also 
Bartello \cite{Bartello} demonstrates 
similar behaviour in the geostrophic energy during the process of geostrophic adjustment in the 3D Boussinesq system. A 
detailed examination of 
this process is in progress --- at this stage we can only conjecture
that, much like behaviour of unstratified rotating fluids \cite{SW-rot}, at long times it is the near resonant interactions 
that feed energy into the slow modes ($\sigma(\vec{k}) = 0$) via interactions between fast waves 
($\sigma(\vec{k}) > 0$) i.e.\ a slow-fast-fast transfer.  \\

\section{Summary and conclusions}

We have studied the evolution of large-scale density (temperature) perturbations in the two-dimensional, stably stratified
Boussinesq equations. The advantage of starting from a state of rest is the observation of various stages through which the
system naturally evolves as governed by the Froude and Reynolds numbers. Starting from a smooth profile, we immediately 
observe the formation
of sharp fronts resulting in a
frontally dominated $k^{-1}$ potential energy spectrum. 
Further, the fronts spontaneously evolve into
highly convoluted sheets accompanied by 
a spectral
re-distribution of energy that culminates in the establishment of a 
$k^{-5/3}$ kinetic energy spectrum. 
Given the rapid decay of energy at this stage, the establishment
of the aforementioned scaling is followed by a gradual downward shift in the spectra, 
the scaling becomes less distinct as the
fields become progressively smoother and finally there emerges a large scale, slowly decaying, vertically sheared, almost horizontal mode
wherein most of the energy is trapped in the potential component --- i.e.\ the Pearson-Linden regime. \\

With regards to the vorticity, the early stages of front formation and energy re-distribution are accompanied by a rapid increase
in enstrophy. Then as the kinetic energy scaling is established we see the onset of an almost monotonic decay of enstrophy. In
particular, immediately after the maximum the decay is primarily exponential with a secondary small amplitude modulation. 
Examining the normalized vorticity
field, motivated by an analogous scenario in the decaying passive scalar problem, shows it to be characterized by an invariant
exponential \pdf{}. This almost exponential decay and the associated invariant \pdf{} persists till we enter the Pearson-Linden
regime. Finally, given the strong diffusive influence, deep into the Pearson-Linden regime the \pdf{} relaxes towards a Gaussian profile. \\

An interesting aspect of this problem is the inviscid limit, i.e.\ $\gamma \rightarrow 0$. If the active scalar system
maintains its regularity i.e.\ $|\nabla \theta|,|\nabla u|,|\nabla w| < \infty$, then as $\gamma \rightarrow 0$ we expect 
$\frac{\partial E}{\partial t}=0$.   However, this does not imply $\gamma <(\nabla
\omega)^2> \rightarrow 0$. In fact, given that the flow will not decay, we conjecture that 
the decay of
enstrophy and the associated invariant \pdf{}s will be established but the flow will never enter the Pearson-Linden regime.
Further, the potential energy spectrum is expected to follow a $k^{-1}$ scaling due to the presence of fronts, whereas the
scaling of the kinetic energy is problematic due to a lack of dissipation. In fact, it is quite possible to end up with equipartition
leading to a pile up of kinetic energy in the largest available wavenumbers \cite{Rose}. On the other hand, if the system 
loses its regularity then $\frac{\partial E}{\partial t}<0$ even in the limit $\gamma \rightarrow 0$, and should behave as in
the presently studied situation with fixed $\gamma > 0$. It is worth noting that issues of a similar sort arise in the classical problem of 
geostrophic adjustment in the shallow water equations --- as pointed out by Killworth \cite{Killworth} (see also Kuo \& Polvani \cite{Kuo-Pol}) --- 
wherein the energy deficit of the geostrophic state 
has implications for wave-breaking and the generation of discontinuities from initially smooth data. In the present 
context it would be interesting to see how the difference in energy between the initial and final states scales with 
diffusivity. \\

An obvious extension of the present work is the consideration of the fully three-dimensional (and also possibly rotating) problem --- as
mentioned in the introduction, the issue of balance and the spontaneous generation of imbalanced (or fast) waves is an active area
of work \cite{Ford}. Indeed, the qualitative similarity to certain aspects of decaying
geostrophic adjustment in the rotating stratified three-dimensional
Boussinesq system --- elegantly elucidated by Bartello \cite{Bartello} --- further motivates such an endevour. \\

\clearpage

\begin{center}
List of Figures 
\end{center}
\begin{itemize}

\item Figure 1 : Upper Panel : Potential, kinetic and total energy with time. Lower Panel : Enstrophy as a
function of time.
The simulation is carried out at $750 \times 750$ resolution with $\gamma=5 \times 10^{-4}$. The other parameters
are specificed in the main text. Note that the simulation was carried out till $t \approx 175$s, this
plot only extends upto $t \approx 60$s so as not to be dominated by the Pearson-Linden regime.

\item Figure 2 : Same as Fig.\ (\ref{figure1}) but the different stages are split up.
Upper Panel : Potential, kinetic and total energy with time in the different stages. The first column shows the relatively inviscid
evolution (till about $t \approx 2$s) followed by the generation of small scale structures. The second column
starts with well defined spectral scaling of the KE and PE and signals the onset of enstrophy decay.
The third column shows the slowly decaying Pearson-Linden regime where most of the energy
is in the potential component. Lower Panel :
Corresponding enstrophy plots.

\item Figure 3 : Snapshots of the temperature field,
with reference to the Fig. (\ref{figure1}) these are at
$t=1.39, 2.34$ seconds respectively.
The emergence of the fronts perpendicular to the ambient stratification is clearly evident. 

\item Figure 4 : 
$\log (<\omega^2>)$ Vs. time that clearly shows a modulated exponential decay immediately after the enstrophy attains its maximum. The initial portion of the
curve for very small times has been omitted for clarity. Note that at long times, as one enters the
Pearson-Linden regime the rate of decay slows and its nature is quite different.

\item Figure 5 :  Snapshots of the vorticity field during the modulated exponential
decay of enstrophy. Note the
field consists of blobs of vorticity separated by ridges of intense enstrophy dissipation.  We encourage the reader to view these
images in color as a grey scale printout masks the sharp features.

\item Figure 6 : \pdf{}s of the normalized vorticity field. The upper panel shows the approach to a self-similar profile, note
the decrease in intermittency with time.
The two bunches of curves in the lower panel consists of
profiles evenly spanning $t\in[12,28]$ sec and $t\in[126,173]$ sec respectively. The upper bunch represents the interval during
which the enstrophy decay is exponential in character (Fig. \ref{figure4}) whereas
the lower bunch shows the \pdf{}'s in
the Pearson-Linden regime.

\item Figure 7 : Snapshots of the $\theta-$ field corresponding to Fig. (\ref{figure5}).
Note that even though there exists a background layering, the fronts riding atop this layering are oriented in a fairly 
random manner. In other words, irrespective of direction a one-dimensional cut of the $\theta-$ field will encounter a
frontal jump. 

\item Figure 8 : Spectra of the kinetic and potential energy from 1D cuts of the $\theta,u,w$ fields. The cuts are made in
the vertical and horizontal directions and the scaling is identical in both the cases (these plots are the averages of the two
cases).
The two bunches of curves have been shifted for clarity, the upper bunch are the KE spectra with the dot-dash line showing a
$k^{-5/3}$ spectrum. Similarly the lower bunch are the PE spectra with the corresponding dot-dash line showing a $k^{-1}$ spectrum.
The kinetic energy attains $k^{-5/3}$ scaling by about
$t \approx 7$s. This is clearly seen in the compensated spectrum plotted at $t=6.7$s, 
after which we see the the gradual downward shift of the spectra, the extension of the diffusive roll-off to
larger scales and the scaling becomes progressively less distinct. 
Specifically, the plots are evenly spaced in the interval
$t \in [6,10]$s.

\item Figure 9 : The establishment of a quasi-horizontal temperature field as we enter the final stages of
decay --- i.e.\ a Pearson-Linden regime.
Referring to
Fig.\ (\ref{figure1}) we see that almost all the energy in the system is now in the potential component.

\end{itemize}


\begin{thebibliography}{99}

\bibitem{Gill} A. Gill, {\em Atmosphere-Ocean Dynamics},
Academic Press, International Geophysics Series, Vol. 30 1982.

\bibitem{RL} J. Riley and M.-P. Lelong, "Fluid motions in the presence of strong stable stratification,"
Annu. Rev. of Fluid Mech. {\bf 32}, 613 (2000).

\bibitem{RMW} J. Riley, R. Metcalfe and M. Weissman , "Direct numerical simulations of homogeneous turbulence in density-
stratified fluids," in {\it Nonlinear Properties of Internal Waves},
AIP Conference Proceedings, ed. B. West, 79 (1981).

\bibitem{Herring} O. Metais and J. Herring, "Numerical simulations of freely evolving turbulence in stably stratified fluids,"
J. Fluid. Mech. {\bf 202}, 117 (1989).

\bibitem{Godeferd} F. Godeferd and C. Cambon, "Detailed investigation of energy transfers in homogeneous stratified turbulence,"
Phys. of Fluids {\bf 6}, 2084 (1994).

\bibitem{SW} L.M. Smith and F. Waleffe, "Generation of slow large-scales in forced rotating stratified turbulence,"
J. Fluid. Mech. {\bf 451}, 145 (2002).

\bibitem{Majda-book} A. Majda, {\em Introduction to PDEs and Waves for the Atmosphere and Ocean},
American Mathematical Society (2003).

\bibitem{LR} M.-P. Lelong and J. Riley, "Internal wave-vortical mode interactions in strongly stratified flows,"
J. Fluid. Mech. {\bf 232}, 1 (1991).

\bibitem{Ford} R. Ford, M. McIntyre and W. Norton, "Balance and the slow quasi-manifold : some explicit results,"
J. Atmos. Sci. {\bf 57}, 1236 (2000).

\bibitem{BS} P. Bouruet-Aubertot, J. Sommeria J and C. Staquet, "Breaking of standing internal gravity waves through 
two-dimensional instabilities,"
J. Fluid. Mech. {\bf 285}, 265 (1995).

\bibitem{BS1} P. Bouruet-Aubertot, J. Sommeria J and C. Staquet, "Stratified turbulence produced by 
internal wave breaking : two-dimensional numerical experiments,"
Dyn. Ocean Atmos. {\bf 23}, 357 (1996).

\bibitem{Orlanski} I. Orlanski and C. Cerasoli, "Resonant and non-resonant wave-wave interactions for internal gravity waves," in 
{\it Marine Turbulence},
Elsevier Oceanography Series 28, ed. J. Nihoul, 65 (1980).

\bibitem{S-rev} C. Staquet and J. Sommeria, "Internal gravity waves: From instabilities to turbulence,"
Annu. Rev. Fluid Mech. {\bf 34}, 559 (2002).

\bibitem{Celani-active} A. Celani, M. Cencini, A. Mazzino and M. Vergassola, "Active and passive fields face to face,"
New Journal of Physics, {\bf 6}, Art. No. 72 (2004).

\bibitem{Jai-chaos} J. Sukhatme and R.T. Pierrehumbert, "Surface quasi-geostrophic turbulence : the study of an active scalar,"
Chaos,{\bf 12}, 439 (2002).

\bibitem{TS} S. Toh and E. Suzuki, "Entropy cascade and Energy Inverse transfer in two-dimensional convective turbulence,"
Phys. Rev. Lett. {\bf 73}, 1501 (1994).

\bibitem{Celani1} A. Celani, A. Mazzino and M. Vergassola, "Thermal plume turbulence,"
Phys. of Fluids {\bf 13}, 2133 (2001).

\bibitem{Celani2} A. Celani, T. Matsumoto, A. Mazzino and M. Vergassola, "Scaling and universality in turbulent convection,"
Phys. Rev. Lett. {\bf 88}, 054503 (2002).

\bibitem{Cordoba} D. Cordoba and C. Fefferman, "Scalars convected by two-dimensional incompressible flow,"
Comm. Pure and Appl. Math. {\bf LV}, 255 (2002).

\bibitem{Pearson} H. Pearson and P. Linden, "The final stage of decay of turbulence in stably stratified fluid,"
J. Fluid. Mech. {\bf 134}, 195 (1983).

\bibitem{Shepherd} T. Shepherd, "A unified theory of available potential energy,"
Atmos-Ocean, {\bf 31}, 1 (1993).

\bibitem{Kraichnan-Mont} R. Kraichnan and D. Montgomery, "Two-dimensional turbulence,"
Rep. Prog. Phys. {\bf 43}, 35 (1980).

\bibitem{Phil_1} O. Phillips, "The generation of clear-air turbulence by the degradation of internal waves,"
in {\it Atmospheric Turbulence and Radio Wave Propagation}, Nauka Moscow, eds. A. Yaglom and V. Tartarski, 130, 1965.

\bibitem{Vassi} J. Vassilicos and J. Hunt, "Fractal dimensions and spectra of interfaces with application to turbulence,"
Proc. R. Soc. Lond. A {\bf 435}, 505 (1991).

\bibitem{Ray-fractal} R.T. Pierrehumbert, "Spectra of tracer distributions : A geometric approach," in 
{\it Nonlinear Phenomena in Atmospheric and Oceanic Sciences},
The IMA Volumes in Mathematics and its Applications, eds. G. Carnevale and R.T. Pierrehumbert, 27 (1992).


\bibitem{Phil_2} O. Phillips, "On the Bolgiano and Lumley-Schur theories of the bouyancy subrange,"
in {\it Atmospheric Turbulence and Radio Wave Propagation}, Nauka Moscow, eds. A. Yaglom and V. Tartarski, 121, 1965.

\bibitem{Lumley} J. Lumley, "The spectrum of nearly inertial turbulence in a stably stratified fluid,"
J. Atmos. Sci. {\bf 21}, 99 (1964).

\bibitem{Sinai-Yakhot} Ya. Sinai and V. Yakhot, "Limiting probability distributions of a passive scalar in a random velocity field,"
Phys. Rev. Lett. {\bf 63}, 1962 (1989).

\bibitem{Jai-1} J. Sukhatme and R.T. Pierrehumbert, "Decay of Passive Scalars Under the Action of Single Scale
Smooth Velocity Fields in Bounded 2D Domains : From  non self similar PDFs to self similar eigenmodes,"
Phys. Rev. E {\bf 66}, 056302 (2002).

\bibitem{Jai-2} J. Sukhatme, "Probability density functions of decaying passive scalars in periodic domains :
an application of Sinai-Yakhot theory,"
Phys. Rev. E {\bf 69}, 056302 (2004).

\bibitem{Yakhot-etal} V. Yakhot, S. Orszag, S. Balachandar, E. Jackson, Z.-S. She and L. Sirovich, "Phenomenological theory
of probability distributions in turbulence,"
J. of Scientific Computing {\bf 5}, 199 (1990).

\bibitem{Bartello} P. Bartello, "Geostrophic adjustment and inverse cascades in rotating stratified turbulence,"
J. Atmos. Sci. {\bf 52}, 4410 (1995).

\bibitem{Smith-short} L.M. Smith, "Numerical study of two-dimensional stratified turbulence,"
Contemporary Mathematics {\bf 283}, 91 (2001).

\bibitem{SW-rot} L.M. Smith and F. Waleffe, "Transfer of energy to two-dimensional large scales in forced, rotating
three-dimensional turbulence,"
Phys. of Fluids {\bf 11}, 1608 (1999).

\bibitem{Rose} H. Rose and P. Sulem, "Fully developed turbulence and statistical mechanics,"
Journal de Physique {\bf 39}, 441 (1978).

\bibitem{Killworth} P. Killworth, "The time-dependent collapse of a rotating fluid cylinder,"
J. Phys. Oceanogr. {\bf 22}, 390 (1992).

\bibitem{Kuo-Pol} A. Kuo and L. Polvani, "Time-dependent fully nonlinear geostrophic adjustment,"
J. Phys. Oceanogr. {\bf 27}, 1614 (1997).

\end{thebibliography}
\end{document}